\documentclass[twocolumn]{jpsj3}
\usepackage{graphicx}
\usepackage{color}
\usepackage{bm}

\usepackage{url}

\title{
Probability Conservation and Localization
in a One-Dimensional Non-Hermitian System
}

\author{
Yositake Takane$^{1}$, Shion Kobayashi$^{1}$, and Ken-Ichiro Imura$^{2}$
}

\inst{
$^{1}$Graduate School of Advanced Science and Engineering,\\
Hiroshima University, Higashihiroshima, Hiroshima 739-8530, Japan\\
$^{2}$Institute of Industrial Science, The University of Tokyo, 277-8574, Japan
}

\recdate{ \hspace{50mm} }

\abst{
We consider transport through a non-Hermitian conductor
connected to a pair of Hermitian leads
and analyze the underlying non-Hermitian scattering problem.
In a typical non-Hermitian system,
such as a Hatano--Nelson-type asymmetric hopping model,
the continuity of probability and probability current
is broken at a local level.
As a result, the notion of transmission and reflection probabilities
becomes ill-defined.
Instead of these probabilities,
we introduce the injection rate $R_{\rm I}=1-|{\cal R}|^2$
and the transmission rate $R_{\rm T}=|{\cal T}|^2$
as relevant physical quantities, where ${\cal T}$ and ${\cal R}$ are
the transmission and reflection amplitudes, respectively.
In a generic non-Hermitian case,
$R_{\rm I}$ and $R_{\rm T}$ have independent information.
We provide a modified continuity equation in terms of
incoming and outgoing currents, from which we derive
a global probability conservation law that relates $R_{\rm I}$ and $R_{\rm T}$.
We have tested the usefulness of our probability conservation law
in the interpretation of numerical results for
non-Hermitian localization and delocalization phenomena.
}
 
\begin{document}
\maketitle

\section{Introduction}

The non-Hermitian quantum mechanics prescribed by a non-Hermitian Hamiltonian
is relevant to the description of an open quantum system,
i.e., a quantum system coupled to an environment.~\cite{open_q,open_q2}
Historically, the idea of non-Hermitian quantum mechanics
dates back to that of an optical potential~\cite{opt1,opt2,opt3}
in the scattering theory for the description of nuclear decay
in terms of resonant states with complex eigenvalues.
In Hermitian quantum mechanics, the scattering problem
is a typical setup and describes a situation in which
an incident wave is scattered in various directions by a given potential.
Let us consider a one-dimensional lattice system of infinite length
with the lattice constant $a$
and assume that a finite region of $n\in [1,N]$ serves as a scattering region,
where $n$ specifies a site on the lattice system.
When a plane wave $e^{ikna}$ ($n \le 0$) is incident in the scattering region,
it is either transmitted or reflected.
The corresponding wave function is given by
$e^{ikna}\ + {\cal R}e^{-ikna}$
in the region of $n \le 0$
and by ${\cal T}e^{ikna}$ in the region of $n \ge N+1$,
where ${\cal T}$ and ${\cal R}$ are
the transmission and reflection amplitudes, respectively.
All the information of the scattering problem is encoded
in these complex amplitudes.
The quantities $T = |{\cal T}|^2$ and $R = |{\cal R}|^2$
satisfy the identity
\begin{equation}
\label{ide}
  T+R = 1,
\end{equation}
which is interpreted as the manifestation of
probability conservation.
This allows us to interpret $T$ and $R$
as the transmission and reflection probabilities, respectively.
In Hermitian quantum mechanics,
identity~(\ref{ide}) always holds.
However, Eq. (\ref{ide}) does not necessarily hold in a non-Hermitian
scattering problem.~\cite{andrianov,levani,deb,muga,cannata,jones,znojil,reso1,
jin,abhinav,reso2,kalozoumis,garmon,zhu,ruschhaupt,burke,shobe}
As a result, $T$ and $R$ cannot be regarded as probabilities.
They need to be reinterpreted.

More generically, setting aside the scattering problem for the moment,
the probability conservation in quantum mechanics can be expressed locally
in the form of a continuity equation of probability and probability current.
In a one-dimensional lattice system, this reads
\begin{equation}
 \frac{\partial\rho_{n} (t)}{\partial t}
 =  j_{n-\frac{1}{2}}(t) -j_{n+\frac{1}{2}}(t),
\label{conti}
\end{equation}
where the probability $\rho_n(t)$ of finding an electron at the $n$th site
is related to the corresponding wave function $\psi_n(t)$
as $\rho_n(t)=|\psi_n(t)|^2$.~\cite{comment1}
Note that the probability $\rho_n(t)$ is defined on the $n$th site,
whereas the probability current $j_{n+\frac{1}{2}}(t)$ is
defined on the link $(n,n+1)$.
The explicit form of $j_{n+\frac{1}{2}}(t)$ is model-dependent
and is determined by the quantum dynamics of the system
driven by the Schr\"odinger equation.
In the case of the simple tight-binding model
\begin{equation}
\label{H0}
 H_0 = \sum_{n}
       \biggl[ -\Gamma |n\rangle\langle n+1| -\Gamma |n+1 \rangle\langle n|
       \biggr],
\end{equation}
$j_{n+\frac{1}{2}}(t)$ reads
\begin{equation}
\label{j0}
  j_{n+\frac{1}{2}}(t)
   = i\frac{\Gamma}{\hbar}
     \left[  \psi_{n+1}(t)^{*}\psi_n(t)
           - \psi_{n+1}(t)\psi_n(t)^{*} \right].
\end{equation}
The continuity equation~(\ref{conti}) assures
a consistent probabilistic interpretation of quantum mechanics.
In a non-Hermitian system, the continuity equation~(\ref{conti})
does not hold as it is.~\cite{abhinav,cannata,bagchi,NH2019TD}
Let us consider the simplest case where an imaginary scalar potential
$i\gamma_n$ is added to Eq.~(\ref{H0}) as
\begin{equation}
\label{H_GL}
  H_{\gamma}
    = \sum_{n}
      \biggl[ - \Gamma |n\rangle\langle n+1| - \Gamma |n+1 \rangle\langle n|
             + i\gamma_n |n\rangle\langle n|
      \biggr].
\end{equation}
In this case, the continuity equation~(\ref{conti}) acquires a correction:
\begin{equation}
\label{conti2}
 \frac{\partial\rho_n(t)}{\partial t}
 = j_{n-\frac{1}{2}}(t) - j_{n+\frac{1}{2}}(t)
   +\frac{2}{\hbar} \gamma_{n} \rho_{n}(t),
\end{equation}
revealing that the probability conservation is broken
in the sense of Hermitian quantum mechanics.
The imaginary scalar potential $i\gamma_n$ serves as a local source
of gain or loss of the probability.
One can regard that this describes the injection or leakage of
the probability current due to the coupling of the system with a reservoir.
If this hypothetical reservoir is taken into consideration,
one can say that the probability is still conserved.
In the sense of this modified probabilistic interpretation,
one can interpret the wave function $\psi_n(t)$
as a complex probability amplitude.
However, in a more generic non-Hermitian situation,
it is unclear whether such an {\it ad hoc} reinterpretation
of the breaking of probability conservation is always possible.
For example, in the case of the Hatano--Nelson-type tight-binding model
with asymmetric hopping,~\cite{HN1,HN2}
\begin{equation}
   \label{HN-asymmetric}
  H_{\rm HN}
  = \sum_{n} \biggl[ - \Gamma_L |n \rangle \langle n+1| 
                    - \Gamma_R |n+1 \rangle\langle n| \biggr],
\end{equation}
such a reinterpretation as that in the case of Eq.~(\ref{conti2})
is ineffective.~\cite{NH2019TD,KK}

In this paper, we attempt to develop a more general framework to overcome
the apparent breakdown of probability conservation in a non-Hermitian system.
As a concrete example, we focus on the one-dimensional scattering problem
in which a scattering region is described by a non-Hermitian Hamiltonian
with both the asymmetric hopping and the imaginary scalar potential.
For this system, we obtain a global probability conservation law
that leads us to define the injection rate $R_{\rm I}=1-|{\cal R}|^2$
and the transmission rate $R_{\rm T}=|{\cal T}|^2$
as relevant physical quantities.
We show that the probability conservation law is useful in interpreting
numerical results for the localization and delocalization phenomena
in the system.

\begin{figure}[btp]
\begin{center}
\includegraphics[height=2.0cm]{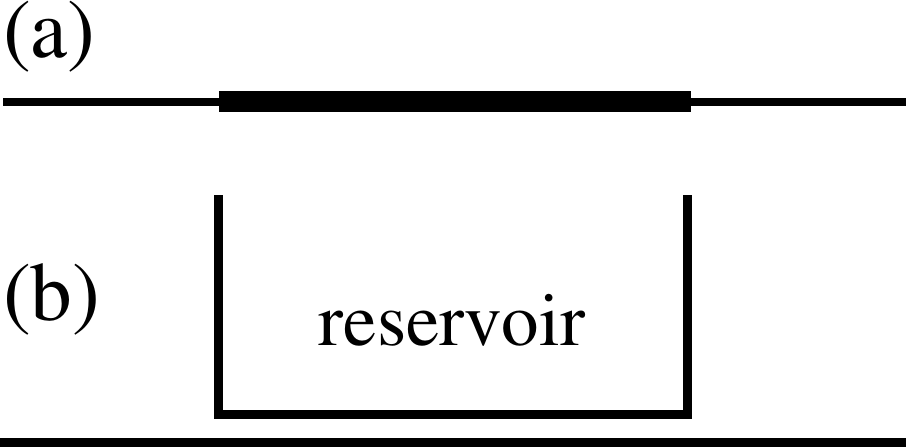}
\end{center}
\caption{
Schematics of the system.
(a) One-dimensional non-Hermitian model consisting of a non-Hermitian region
(thick solid line) and two Hermitian leads (solid lines).
(b) Hermitian system consisting of a one-dimensional system and a reservoir,
which underlies the non-Hermitian model shown in panel (a).
The non-Hermiticity of the model in panel (a) is assumed to arise from
the coupling of the one-dimensional system with the reservoir.
}
\end{figure}

\section{Non-Hermitian scattering problem}

We consider the following setup extended to one-dimensional lattice sites
$n=0,\pm1,\pm 2,\cdots$ in which a non-Hermitian scattering region
$n \in [1,N]$ prescribed by a non-Hermitian Hamiltonian $H_{\rm S}$
is connected to two semi-infinite Hermitian leads.
The left lead extended to $n \rightarrow -\infty$ is described by $H_{\rm L}$,
and the right lead extended to $n \rightarrow \infty$
is described by $H_{\rm R}$.
The total Hamiltonian is given by
$H = H_{\rm L}+H_{\rm S}+H_{\rm R}$,
where
\begin{align}
   H_{\rm L}
 & = \sum_{n \le 0}
     \biggl[ - \Gamma |n \rangle \langle n+1| - \Gamma |n+1 \rangle \langle n|
     \biggr],
       \label{H_L}
         \\
   H_{\rm S}
 & = \sum_{1 \le n \le N-1}
     \biggl[ - \Gamma_{\rm L} |n \rangle \langle n+1| 
             - \Gamma_{\rm R} |n+1 \rangle \langle n|
     \biggr]
   \nonumber \\
 & \hspace{10mm}
     + \sum_{1 \le n \le N} \left(V_{n}+i\gamma_{n}\right)
                            |n \rangle \langle n|,
       \label{H_S}
   \\
   H_{\rm R}
 & = \sum_{n \ge N}
     \biggl[ - \Gamma |n \rangle \langle n+1| - \Gamma |n+1 \rangle \langle n|
     \biggr],
       \label{H_R}
\end{align}
where $\Gamma$, $\Gamma_{\rm L}$, and $\Gamma_{\rm R}$ as well as
$V_{n}$ and $\gamma_{n}$  with $n \in [1,N]$ are chosen to be real.
Note that $H_{\rm S}$ contains two sources of non-Hermiticity:
\begin{itemize}
\item 
the imaginary scaler potential $i \gamma_{n}$ representing gain or loss,
\item
the asymmetric hopping $\Gamma_{\rm L} \neq \Gamma_{\rm R}$,
\end{itemize}
and the random scattering potential $V_{n}$ as well.
The advantage of this setup is that we can set the energy of
a stationary scattering state to be a real value.
Let us assume that a plane wave $\psi_n^{\rm (in)} = e^{ikna}$ ($n\le 0$)
specified by a real wavenumber $k$ is incident in
the non-Hermitian scattering region $n\in [1,N]$ from the left lead.
Inside the left lead, its energy is fixed to be a real value of
$E = -2\Gamma \cos(ka)$;
hence, the energy of the resulting stationary scattering state
$|\psi\rangle$ must also have the same real energy everywhere in the system.
The stationary scattering state
\begin{equation}
|\psi(t)\rangle = e^{-i\frac{Et}{\hbar}} \sum_n \psi_n |n\rangle,
\label{psi_0}
\end{equation}
with
\begin{align}
   \psi_n = e^{ikna} + {\cal R} e^{-ikna}
\label{psi_1}
\end{align}
for $n \le 0$ and
\begin{align}
\psi_n = {\cal T} e^{ikna}
\label{psi_2}
\end{align}
for $n \ge N+1$ satisfies the time-dependent Schr\"{o}dinger equation
\begin{equation}
i\hbar\frac{\partial}{\partial t}|\psi(t)\rangle
  = H |\psi(t)\rangle,
\label{Sch}
\end{equation}
where the total Hamiltonian $H$ is composed of the three parts
given in Eqs.~(\ref{H_L})--(\ref{H_R}).

The two panels in Fig.~1 illustrate our setup consisting of
a non-Hermitian conductor connected to two Hermitian leads.
The central scattering region is described by
an effective non-Hermitian Hamiltonian [see panel~(a)].
We assume that its non-Hermiticity arises from the coupling of
an underlying Hermitian system to an external reservoir [see panel~(b)].
However, we do not attempt to derive the effective Hamiltonian~(\ref{H_S})
from a microscopic Hermitian model.~\cite{gong,mcdonald,liu}

As for alternative realizations of a Hatano--Nelson-type
asymmetric hopping model, we refer the readers to
recent experiments,~\cite{HN_exp1,HN_exp2,HN_exp3,HN_exp4}
in which the so-called non-Hermitian skin effect characteristic of
such an asymmetric hopping model under the open boundary condition
has been demonstrated.

\begin{figure}[btp]
\begin{center}
\includegraphics[height=1.7cm]{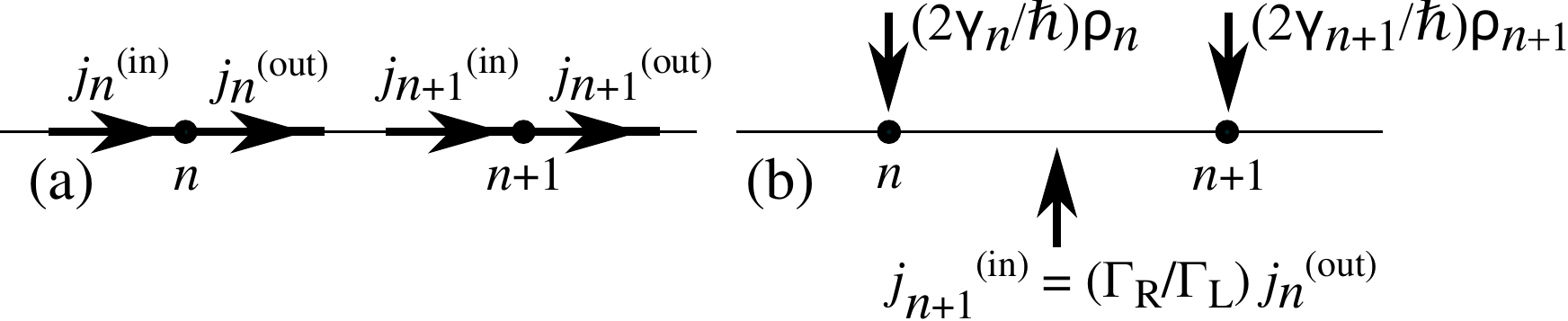}
\end{center}
\caption{
Schematic representation of probability current.
(a) $j_{n}^{\rm (in)}$ and $j_{n}^{\rm (out)}$ are defined for the $n$th site.
(b) $j_{n+1}^{\rm (in)}=\frac{\Gamma_{\rm R}}{\Gamma_{\rm L}}j_{n}^{\rm (out)}$
means that the coupling with the reservoir amplifies (attenuates)
a right-going (left-going) wave
when $\frac{\Gamma_{\rm R}}{\Gamma_{\rm L}} > 1$.
}
\end{figure}

\section{Incoming and outgoing currents}

We have pointed out that in the non-Hermitian scattering region $n\in [1,N]$
prescribed by Eq.~(\ref{H_S}),
the continuity equation~(\ref{conti2}),
associated with the standard Hermitian current $j_{n+\frac{1}{2}}$
given in Eq.~(\ref{j0}), is no longer valid.
Yet, we have also noted that the definition of current that appears
in the continuity equation is generally model-dependent.
If so, can we think of an alternative definition of current that makes
the continuity equation valid in the non-Hermitian scattering region?

To find such an expression of current, we consider the time evolution of
an arbitrary state,
\begin{align}
    \label{eq:arb-state}
  |\psi (t)\rangle = \sum_{n}\psi_n(t)| n\rangle ,
\end{align}
driven by the Schr\"{o}dinger equation~(\ref{Sch}).
Its Hermitian conjugate $\langle\psi(t)| = |\psi(t)\rangle^{\dagger}$ obeys
\begin{equation}
 -i\hbar\frac{\partial}{\partial t}\langle\psi(t)|
 = \langle\psi(t)| H^{\dagger},
\label{Sch*}
\end{equation}
where $H$ is generically non-Hermitian (i.e., $H \neq H^{\dagger}$).
By considering the time evolution of the probability,
\begin{align}
   \rho_n(t)= \langle\psi(t)|n\rangle \langle n|\psi(t)\rangle,
\label{rho_t}
\end{align}
one finds that the modified continuity equation
\begin{align}
     \label{conti_mod}
   \frac{\partial}{\partial t}\rho_n(t)
   = j_n^{\rm (in)}(t) - j_n^{\rm (out)}(t)
     + \frac{2\gamma_{n}}{\hbar}\rho_n(t)
\end{align}
holds in the scattering region
if one introduces the incoming current $j_n^{\rm (in)}(t)$
and the outgoing current $j_n^{\rm (out)}(t)$ defined as
\begin{align}
     \label{j_in}
   j_n^{\rm (in)}(t)
   & = i\frac{\Gamma_{\rm R}}{\hbar}
       \left[\psi_n(t)^{*}\psi_{n-1}(t)-\psi_n(t)\psi_{n-1}(t)^{*}\right] ,
     \\
\label{j_out}
   j_n^{\rm (out)}(t)
   & = i\frac{\Gamma_{\rm L}}{\hbar}
       \left[\psi_{n+1}(t)^{*}\psi_n(t)-\psi_{n+1}(t)\psi_n(t)^{*}\right] .
\end{align}
As shown in Fig.~2(a), the two probability currents
$j_n^{\rm (out)}(t)$ and $j_{n+1}^{\rm (in)}(t)$
are associated with the same link $(n,n+1)$.
They differ from each other when $\Gamma_{\rm L} \neq \Gamma_{\rm R}$
and are necessary to describe local probability conservation
in the presence of asymmetric hopping.

In the Hermitian region consisting of the left and right leads,
the incoming current Eq.~(\ref{j_in})
and the outgoing current Eq.~(\ref{j_out})
reduce to the Hermitian current (\ref{j0}).
That is, both $j_n^{\rm (out)}(t)$ and $j_{n+1}^{\rm (in)}(t)$ reduce to
$j_{n+\frac{1}{2}}(t)$ for $n\le 0$ and for $n\ge N$.

\section{Local and global probability conservation}

Let us consider a more specific situation in which $|\psi (t)\rangle$ given in
Eq. (\ref{eq:arb-state}) is replaced with a stationary scattering state
specified by Eq. (\ref{psi_0}) with real energy.
In this case, $\rho_n$, $j_n^{\rm (in)}$, and $j_n^{\rm (out)}$
do not depend on time and Eq.~(\ref{conti_mod}) yields
\begin{align}
     \label{eq:local-conservation}
   j_n^{\rm (out)} = j_n^{\rm (in)} + \frac{2\gamma_{n}}{\hbar}\rho_n.
\end{align}
Note that $j_n^{\rm (out)}$ flowing out of the $n$th site is different from
$j_n^{\rm (in)}$ flowing into the same site.
The difference is given by the second term on the right-hand side of
Eq. (\ref{eq:local-conservation})
and interpreted as the injection or leakage
(depending on the sign of $\gamma_{n}$) 
of the probability current [see Fig.~2(b)]
due to the imaginary potential in Eq.~(\ref{H_S}).
Recall our assumption that this term stems from the coupling of the system
with a reservoir that has been traced out
when deriving the effective Hamiltonian (\ref{H_S}).
The imaginary potential $i\gamma_{n}$ contributes additively to
the probability current
[see Eqs.~(\ref{eq:local-conservation}) and (\ref{glob_cons})].

Let us also focus on $j_{n}^{\rm (out)}$ and $j_{n+1}^{\rm (in)}$,
both of which are associated with the same link $(n,n+1)$ [see Fig.~2(a)].
Comparing the definitions of the incoming and outgoing currents given in
Eqs.~(\ref{j_in}) and (\ref{j_out}), respectively, one immediately finds
\begin{align}
     \label{eq:multiplicative}
   j_{n+1}^{\rm (in)} = \frac{\Gamma_{\rm R}}{\Gamma_{\rm L}} j_n^{\rm (out)}.
\end{align}
This makes the role of asymmetry between $\Gamma_{\rm L}$ and $\Gamma_{\rm R}$
apparent.
On a given link $(n,n+1)$, $j_n^{\rm (out)}$ flowing out of the $n$th site
is augmented by the factor $\Gamma_{\rm R}/\Gamma_{\rm L}$
when it gets out of the link and becomes $j_{n+1}^{\rm (in)}$ flowing
into the neighboring $n+1$th site.
In other words, there has been again an inflow or leakage
(depending on the direction of asymmetry)
of current on the link [see Fig.~2(b)].
Similarly to the case of the imaginary potential $i\gamma_{n}$,
we can interpret that this amplification or attenuation of
the probability current stems from
the coupling of the system with the reservoir.
Note also that the asymmetric hopping
modifies the probability current in a multiplicative manner.

To summarize, Eqs.~(\ref{eq:local-conservation}) and (\ref{eq:multiplicative})
constitute the local conservation law of probability
valid in the non-Hermitian scattering region prescribed by Eq.~(\ref{H_S}),
in which the continuity of the probability current
in the Hermitian sense [see Eq.~(\ref{conti})] is no longer applicable.

Using Eqs.~(\ref{eq:local-conservation}) and (\ref{eq:multiplicative})
iteratively, one finds
\begin{align}
     \label{glob_cons}
  j_N^{\rm (out)}
   =   \left(\frac{\Gamma_{\rm R}}{\Gamma_{\rm L}}\right)^{N-1}j_1^{\rm (in)}
     + \sum_{n = 1}^{N}
       \left(\frac{\Gamma_{\rm R}}{\Gamma_{\rm L}}\right)^{N-n}
       \frac{2\gamma_{n}}{\hbar}\rho_n.
\end{align}
If one recalls that the probability current is conserved
in the left and right Hermitian leads, i.e.,
\begin{align}
  j_1^{\rm (in)} & = j_0^{\rm (out)} = j_0^{\rm (in)} = j_{-1}^{\rm (out)}
                   = \cdots \equiv j_{\rm L},
    \label{j_1} \\
  j_N^{\rm (out)}& = j_{N+1}^{\rm (in)} = j_{N+1}^{\rm (out)}
                   = j_{N+2}^{\rm (in)} = \cdots \equiv j_{\rm R}
    \label{j_N}
\end{align}
with
\begin{align}
  j_{\rm L} & = \frac{2\Gamma\sin(ka)}{\hbar} (1-|{\cal R}|^2),
     \label{j_L}
  \\
  j_{\rm R} & = \frac{2\Gamma\sin(ka)}{\hbar} |{\cal T}|^2,
\label{j_R}
\end{align}
and substitutes Eqs.~(\ref{j_1})--(\ref{j_R}) into Eq.~(\ref{glob_cons}),
one can rewrite Eq.~(\ref{glob_cons}) as
\begin{align}
     \label{eq:global-cons2}
|{\cal T}|^2
  &=  \left(\frac{\Gamma_{\rm R}}{\Gamma_{\rm L}}\right)^{N-1} (1-|{\cal R}|^2)
  \nonumber \\
   &+ \sum_{n = 1}^{N} \left(\frac{\Gamma_{\rm R}}{\Gamma_{\rm L}}\right)^{N-n}
                      \frac{\gamma_{n} \rho_n}{\Gamma\sin(ka)} ,
\end{align}
which serves as a global conservation law of probability.
Note that Eq.~(\ref{eq:global-cons2})
still holds in the presence of the disorder potential $V_n$.

In the Hermitian limit of $\Gamma_{\rm R}/\Gamma_{\rm L}=1$ and $\gamma_{n}=0$,
Eq.~(\ref{eq:global-cons2}) reduces to Eq.~(\ref{ide}) with
\begin{equation}
   0 \le |{\cal T}|^2 \le 1, \hspace{5mm} 0 \le |{\cal R}|^2 \le 1.
\end{equation}
In the non-Hermitian case of $\Gamma_{\rm R}/\Gamma_{\rm L}\neq 1$ and/or
$\gamma_{n} \neq 0$,
$|{\cal T}|^2 \ge 0$ and $|{\cal R}|^2 \ge 0$ still hold, whereas both of
$|{\cal T}|^2$ and $|{\cal R}|^2$ can be larger than $1$.
That is, Eq.~(\ref{ide}) no longer holds.
The case of $|{\cal R}|^2 > 1$ may be rare but is possible;
preliminary numerical calculations show that this is indeed the case
(see the first paragraph of Sect.~5).
In contrast, the case of $|{\cal T}|^2 > 1$ is typical
as a result of the amplification effect
due to $\Gamma_{\rm R}/\Gamma_{\rm L}>1$ as well as $\gamma > 0$.
Indeed, it is likely that the transmission probability
exceeds $1$.~\cite{jones,shobe}
Again, under such circumstances, one can presume that an effective current
is injected from the hypothetical reservoir
that has been traced out in the process of
deriving the effective Hamiltonian (\ref{H_S}).

The invalidity of the global probability conservation law~(\ref{ide})
in the generic non-Hermitian case suggested by Eq.~(\ref{eq:global-cons2})
implies that the interpretation of the quantities
$T=|{\cal T}|^{2}$ and $R=|{\cal R}|^{2}$
as the transmission and reflection probabilities, respectively,
is no longer appropriate.
In the Hermitian case, the Landauer formula~\cite{landauer,buttiker}
relates the two-terminal conductance $G_2$ of a mesoscopic conductor
to the transmission probability $T=|{\cal T}|^2$ as $G_2 = (e^2/h) T$.
To be precise, zero temperature is assumed and
$T$ should be evaluated at the Fermi energy.
For a bias voltage $V$ applied between the left and right leads,
an influx $I_L^+ = (e^2/h)V$ of electrons is injected from the left lead,
whereas the outflux to the right lead is $I_R^+ = T(e^2/h)V$.
The remaining flux, $I_L^- = R(e^2/h)V$, is reflected back to the left lead,
where $R = 1-T$ is the probability that an electron injected from the left lead is reflected back to the same lead.
These ensure that the net current $I$,
\begin{equation}
I = I_L^+ - I_L^- = I_R^+ = T(e^2/h)V ,
\label{net}
\end{equation}
is conserved throughout in the system.
The two-terminal Landauer formula~\cite{landauer} describes such a situation.
In the non-Hermitian case of invalidated probability conservation
with typically $T > 1$ and possibly $R > 1$,
the central identity of Eq.~(\ref{net}) fails and
Eq.~(\ref{ide}) should be replaced with Eq.~(\ref{eq:global-cons2}).

The net currents $I_L = I_L^+ - I_L^-$ on the left incident side
and $I_R = I_R^+$ on the right transmitted side differ
and have independent information.
In this case, one should deal with $I_L$ and $I_R$ independently
by introducing the following two quantities:
\begin{align}
   \label{R_I}
 I_L/I_L^+ & = 1-R \equiv R_{\rm I}, 
     \\
   \label{R_T}
 I_R/I_L^+ & = T \equiv R_{\rm T}.
\end{align}
Keeping in mind that $T$ and $R$ should no longer be called
probabilities,
let us call $R_{\rm I}$ and $R_{\rm T}$
the injection and transmission rates, respectively.

The fact that $I_L\neq I_R$ in the generic non-Hermitian case
is also consistent with our previous assumption that an effective current is
either injected from or leaked to a hypothetical reservoir
in the non-Hermitian scattering region.
Such a hypothetical reservoir is assumed to be connected to
the scattering region in the underlying Hermitian model.
In this sense, the non-Hermitian setup considered in this work is certainly
beyond the description based on the two-terminal Landauer conductance $G_2$.

\section{Numerical tests}

Let us consider the injection and transmission rates
introduced in Eqs. (\ref{R_I}) and (\ref{R_T}), respectively.
In the Hermitian limit, they reduce to $R_{\rm I} = R_{\rm T}= T$.
They obviously satisfy $R_{\rm I} \le 1$ and $R_{\rm T} \ge 0$.
However, $R_{\rm I} \ge 0$ and $R_{\rm T} \le 1$ are not guaranteed.
Indeed, preliminary numerical calculations show that $R_{\rm I}$ can become
negative when $\gamma > 0$.
Here and hereafter, we assume for simplicity that the imaginary potential
$i\gamma_{n}$ is uniform in the scattering region of $n\in [1,N]$ as
\begin{align}
  i\gamma_{n} = i\gamma .
\end{align}
The situation $R_{\rm I} < 0$ is induced by injection from
the hypothetical reservoir near the left end of the scattering region
when a large portion of this injection is converted into the left-going current
by a strong disorder such that $R = 1 - R_{\rm I} > 1$.
We leave this situation for a future study
and restrict our consideration to the case of $\gamma \le 0$,
in which $0 \le R_{\rm I} \le 1$.
Once $0 \le R_{\rm I} \le 1$ is ensured,
we accept the situation $R_{\rm T} > 1$
as a result of the amplification due to the non-Hermiticity
describing the coupling with the reservoir.

Let us now examine the localization and delocalization phenomena
in the non-Hermitian disordered system.~\cite{HN1,HN2}
The asymmetry in $\Gamma_{\rm R}$ and $\Gamma_{\rm L}$ is set as
\begin{align}
  \Gamma_{\rm R} & = e^{g}\Gamma ,
     \\
  \Gamma_{\rm L} & = e^{-g}\Gamma .
\end{align}
We calculate $-\langle \log R_{\rm T}\rangle$
and $-\langle \log R_{\rm I}\rangle$ averaged over many samples
with different disorder potentials for
systems of size $N$ ranging from $50$ to $300$.
Here, $-\langle \log R_{\rm T/I}\rangle$ characterizes
the exponential increase or decrease in $R_{\rm T/I}$ with respect to $N$.
Note that $\langle \log R_{\rm I}\rangle = \langle \log R_{\rm T}\rangle$
holds exactly in the Hermitian limit.
The wavenumber is fixed at $k = \frac{\pi}{4a}$.
The sample average is taken as follows.
We assume that $V_{n}$ for each $n \in [1,N]$ is a random number
uniformly distributed in the range of $V_{n} \in[-\frac{W}{2},\frac{W}{2}]$,
where $W$ characterizes the strength of disorder.
Each ensemble average is taken for $10^5$ samples,
except in the two cases, namely,
$(W/\Gamma, g, \gamma/\Gamma) = (0.969, 0.02, 0.0)$ and $(0.9, 0.02, -0.004)$,
in which we have employed $10^6$ samples
to improve the precision of our data.

\begin{figure}[btp]
\begin{tabular}{cc}
\begin{minipage}{0.5\hsize}
\begin{center}
\hspace{-10mm}
\includegraphics[height=3.3cm]{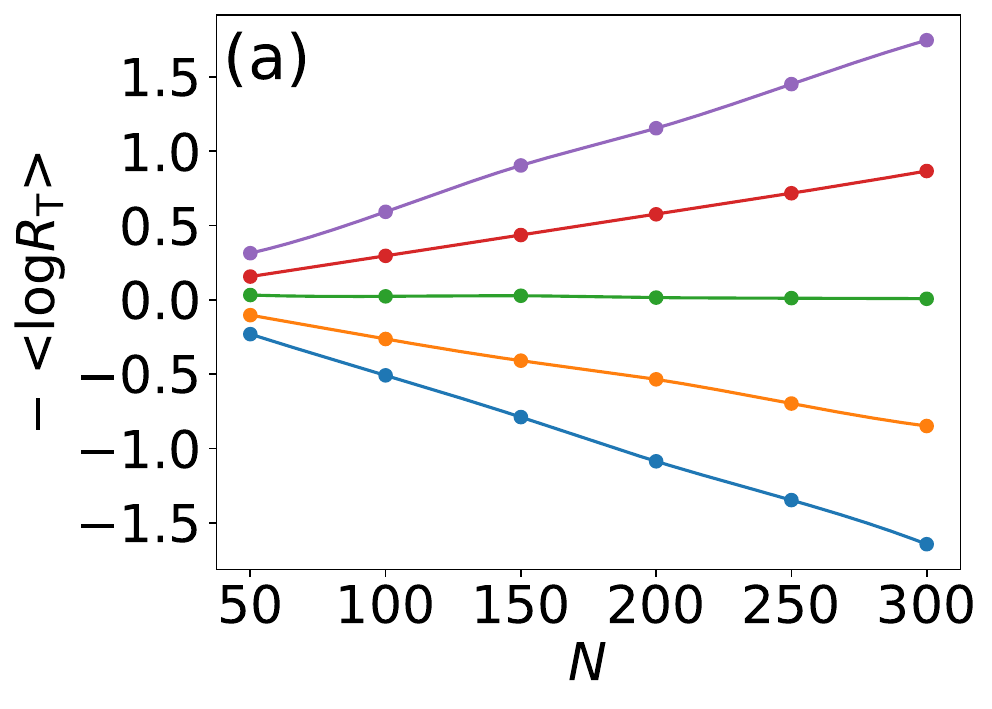}
\end{center}
\end{minipage}
\begin{minipage}{0.5\hsize}
\begin{center}
\hspace{-10mm}
\includegraphics[height=3.3cm]{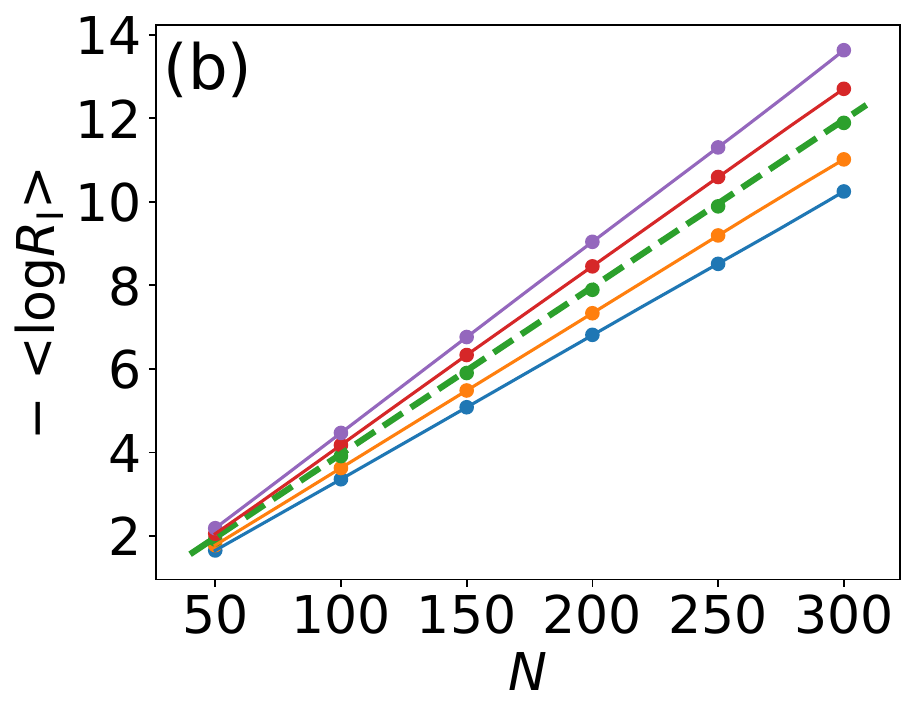}
\end{center}
\end{minipage}
\end{tabular}
\caption{
(Color online)
(a) $-\langle \log R_{\rm T} \rangle$ and (b) $-\langle \log R_{\rm I} \rangle$
in the case of $g = 0.02$ and $\gamma = 0.0$
for $N$ ranging from $50$ to $300$,
where $W/\Gamma = 0.9$ (blue), $0.934$ (orange), $0.969$ (green),
$1.002$ (red), and $1.036$ (violet) from bottom to top.
In panel~(b), a dashed line represents
$-\langle \log R_{\rm I} \rangle = 2g (N-1)$.
Solid lines serve as visual guides.
}
\end{figure}
We first examine the case of $g = 0.02$ and $\gamma = 0$
for $W/\Gamma = 0.9$, $0.934$, $0.969$, $1.002$, and $1.036$.
In the case of $\gamma = 0$,
the global conservation law~(\ref{eq:global-cons2}) is reduced to
\begin{align}
      \label{eq:global-cons_1}
   R_{\rm T} =  e^{2g(N-1)}R_{\rm I} ,
\end{align}
which describes the amplification or attenuation of
the injected probability current from the left lead
owing to the asymmetry in $\Gamma_{\rm L}$ and $\Gamma_{\rm R}$.
Figure~3(a) shows $-\langle \log R_{\rm T}\rangle$ 
as a function of $N$.
A linear relationship between $-\langle \log R_{\rm T}\rangle$ and $N$
with a positive (negative) slope means that $R_{\rm T}$ 
decreases (increases) exponentially with increasing $N$.
We observe from Fig.~3(a) that the slope of $-\langle \log R_{\rm T}\rangle$
changes from a negative value at $W/\Gamma = 0.934$
to a positive value at $W/\Gamma = 1.002$,
whereas the slope is nearly flat at $W/\Gamma = 0.969$.
This shows that the delocalization--localization transition occurs
at the critical value of $W_{\rm c}/\Gamma \approx 0.969$.
Equation~(\ref{eq:global-cons_1}) indicates that
if $-\langle \log R_{\rm T}\rangle$ is nearly flat,
$-\langle \log R_{\rm I}\rangle$ increases with $N$ as
$-\langle \log R_{\rm I}\rangle \approx 2g(N-1)$.
This is consistent with the behavior of $-\langle \log R_{\rm I}\rangle$
shown in Fig.~3(b).
Indeed, we observe in Fig.~3(b) that the data in the case of $W/\Gamma = 0.969$
is on the dashed line of $-\langle \log R_{\rm I} \rangle = 2g (N-1)$.
Note that the behavior of $R_{\rm I}$ is identical to that of
the transmission probability $T$ in the Hermitian limit of $g = 0$
as shown in Appendix.

\begin{figure}[tbp]
\begin{tabular}{cc}
\begin{minipage}{0.5\hsize}
\begin{center}
\hspace{-10mm}
\includegraphics[height=3.3cm]{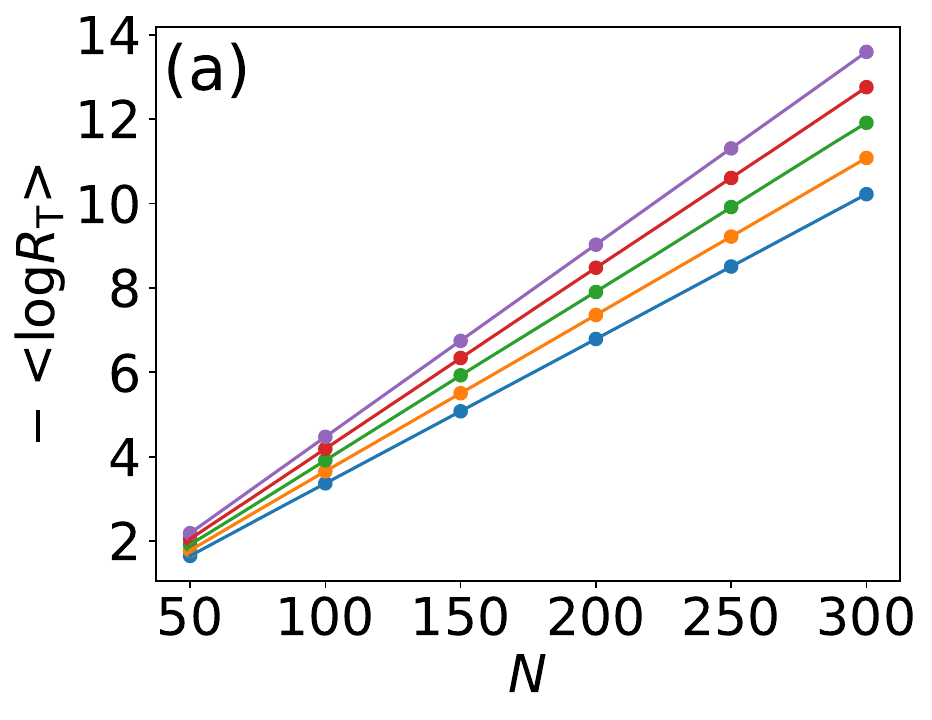}
\end{center}
\end{minipage}
\begin{minipage}{0.5\hsize}
\begin{center}
\hspace{-10mm}
\includegraphics[height=3.3cm]{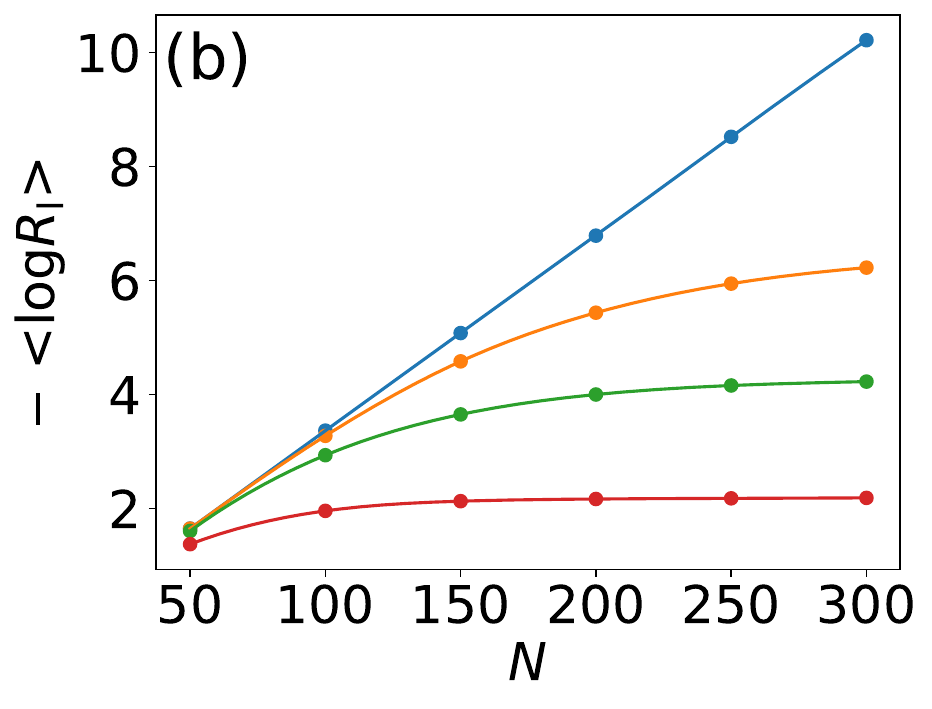}
\end{center}
\end{minipage}
\end{tabular}
\caption{
(Color online)
(a) $-\langle \log R_{\rm T} \rangle$ and (b) $-\langle \log R_{\rm I} \rangle$
in the case of $g = 0.0$ and $W/\Gamma = 0.9$ for $N$
ranging from $50$ to $300$,
where $\gamma/\Gamma = 0.0$ (blue), $-0.002$ (orange), $-0.004$ (green),
$-0.006$ (red), and $-0.008$ (violet) from bottom to top in panel~(a),
and $\gamma/\Gamma = 0.0$ (blue), $-0.00001$ (orange), $-0.0001$ (green),
and $-0.001$ (red) from top to bottom in panel~(b).
Solid lines serve as visual guides.
}
\end{figure}
We next examine the case of $g = 0$
at $W/\Gamma = 0.9$ and several values of $\gamma$.
Figure~4(a) shows $-\langle \log R_{\rm T}\rangle$
for $\gamma/\Gamma = 0.0$, $-0.002$, $-0.004$, $-0.006$, and $-0.008$.
We observe that $R_{\rm T}$ decreases exponentially as $N$ increases,
and this exponential decrease becomes more rapid as $|\gamma|$ increases.
Figure~4(b) shows $-\langle \log R_{\rm I}\rangle$ for
$\gamma/\Gamma = 0.0$, $-0.00001$, $-0.0001$, and $-0.001$.
We observe that when $\gamma$ becomes finite ($|\gamma|>0$),
the exponential decrease in $R_{\rm I}$ seen at $\gamma = 0$
tends to stop at a finite value;
$-\langle \log R_{\rm I}\rangle$ appears saturated at a sufficiently large $N$.
Indeed, the behaviors of $R_{\rm T}$ and $R_{\rm I}$
are contrasting in the case of $\gamma \neq 0$.
Unlike in the case of $R_{\rm T}$, the decrease in $R_{\rm I}$,
after an initial exponential decrease, tends to be moderated,
and $R_{\rm I}$ converges to a small but finite value.
This lower limit of $R_{\rm I}$,
which becomes vanishingly small as $\gamma\rightarrow 0$,
increases with $|\gamma|$.
Let us explain this feature in relation to Eq.~(\ref{eq:global-cons2}),
which is rewritten in the present case as
\begin{align}
     \label{eq:global-cons_2}
   R_{\rm T}
  =  R_{\rm I} + \sum_{n = 1}^{N} \frac{\gamma \rho_{n}}{\Gamma\sin(ka)} .
\end{align}
For a sufficiently large $N$, 
$R_{\rm I}$ takes a value close to its lower limit,
whereas $R_{\rm T}$ continues to decrease exponentially.
In this regime, the dependence of $R_{\rm T}$ on $N$ is determined by
the second term in Eq.~(\ref{eq:global-cons_2}),
which represents the total amount of leakage
due to the constant imaginary potential $i\gamma$.
Assuming $\rho_{n} \propto \exp(-\frac{2n}{\xi})$ for $n \gg 1$
with a decay length $\xi$ normalized by $a$,
we approximately express the second term in Eq.~(\ref{eq:global-cons_2})
with $\gamma < 0$ as
\begin{align}
    \label{eq:integral_1}
  \sum_{n = 1}^{N} \frac{\gamma \rho_{n}}{\Gamma\sin(ka)}
  \sim A \exp\left(-\frac{2N}{\xi}\right) -B,
\end{align}
where $A> 0$ is independent of $N$,
whereas $B > 0$ may weakly depend on $N$.
In the Hermitian limit of $\gamma = 0$,
$\xi$ is identified as the localization length.
We can explain the behaviors of $R_{\rm T}$ and $R_{\rm I}$
if $B$ cancels $R_{\rm I}$  (i.e., $B = R_{\rm I}$)
on the right-hand side of Eq.~(\ref{eq:global-cons_2}).
Thus, the first term on the right-hand side of Eq.~(\ref{eq:integral_1})
describes the exponential decrease in $R_{\rm T}$ as
\begin{align}
  R_{\rm T} \sim A \exp\left(-\frac{2N}{\xi}\right).
\end{align}
Since the exponential decrease in $\rho_{n}$ is enhanced by
the leakage due to the constant imaginary potential $i\gamma$,
the decay length $\xi$ decreases with increasing $|\gamma|$.
This explains the behavior of $R_{\rm T}$ shown in Fig.~4(a).
In turn, the leakage can occur only when the total amount of leakage,
which is nearly equal to $B$, is supplied to the system from the left lead
as indicated by $B = R_{\rm I}$.
Hence, the injection rate $R_{\rm I}$ cannot continue to decrease with $N$,
but tends to be bounded from below.
The lower limit of $R_{\rm I}$ 
increases with increasing $|\gamma|$,
since the total amount of leakage increases with $|\gamma|$.
This explains the behavior of $R_{\rm I}$ shown in Fig.~4(b).

\begin{figure}[btp]
\begin{tabular}{cc}
\begin{minipage}{0.5\hsize}
\begin{center}
\hspace{-10mm}
\includegraphics[height=3.3cm]{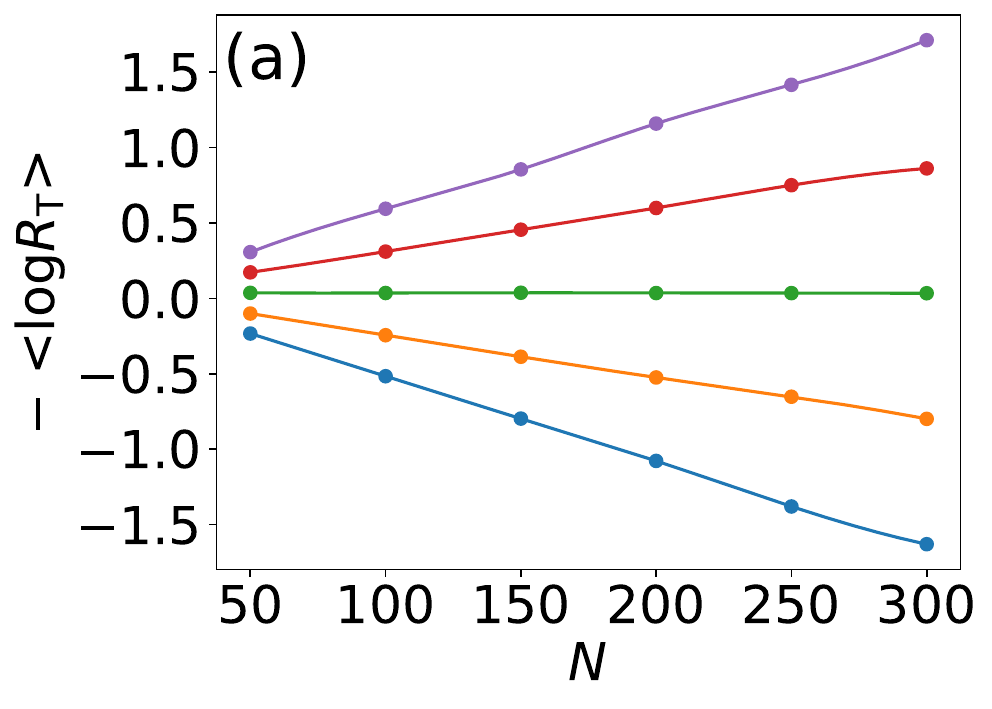}
\end{center}
\end{minipage}
\begin{minipage}{0.5\hsize}
\begin{center}
\hspace{-10mm}
\includegraphics[height=3.3cm]{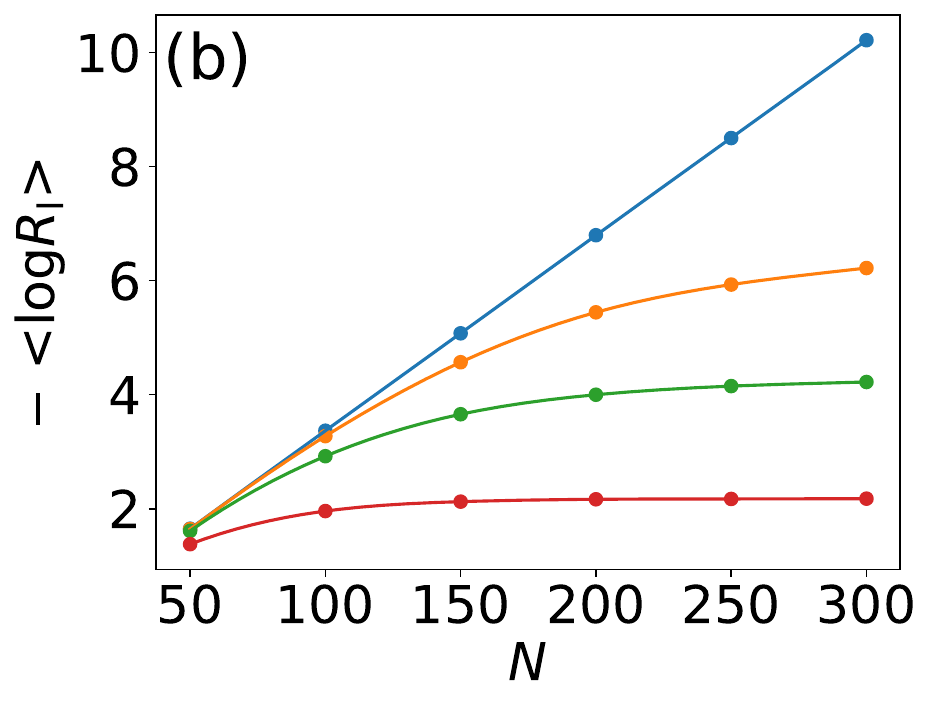}
\end{center}
\end{minipage}
\end{tabular}
\caption{
(Color online)
(a) $-\langle \log R_{\rm T} \rangle$ and (b) $-\langle \log R_{\rm I} \rangle$
in the case of $g = 0.02$ and $W/\Gamma = 0.9$ for $N$
ranging from $50$ to $300$,
where $\gamma/\Gamma = 0.0$ (blue), $-0.002$ (orange), $-0.004$ (green),
$-0.006$ (red), and $-0.008$ (violet) from bottom to top in panel~(a),
and $\gamma/\Gamma = 0.0$ (blue), $-0.00001$ (orange), $-0.0001$ (green),
and $-0.001$ (red) from top to bottom in panel~(b).
Solid lines serve as visual guides.
}
\end{figure}
Finally, we examine the case of $g = 0.02$
at $W/\Gamma = 0.9$ and several values of $\gamma$.
Figure~5(a) shows $-\langle \log R_{\rm T}\rangle$
for $\gamma/\Gamma = 0.0$, $-0.002$, $-0.004$, $-0.006$, and $-0.008$.
We observe from Fig.~5(a) that the slope of $-\langle \log R_{\rm T}\rangle$
changes from a negative value at $\gamma/\Gamma = -0.002$
to a positive value at $\gamma/\Gamma = -0.006$,
whereas the slope is nearly flat at $\gamma/\Gamma = -0.004$.
This shows that the delocalization--localization transition occurs
at the critical value of $\gamma_{\rm c}/\Gamma \approx -0.004$.
We also observe that, for each value of $\gamma$,
the exponential decrease in $R_{\rm T}$ is slower than
that in the case of $g = 0$ shown in Fig.~4(a).
This is again explained by using Eq.~(\ref{eq:global-cons2}),
which is rewritten in this case as
\begin{align}
      \label{eq:global-cons_3}
   R_{\rm T}
  =  e^{2g(N-1)}R_{\rm I}
   + \sum_{n = 1}^{N} \frac{e^{2g(N-n)}\gamma \rho_{n}}{\Gamma\sin(ka)} .
\end{align}
Assuming $\rho_{n} \propto \exp[-2(\frac{1}{\xi}-g)n]$ for $n \gg 1$,
we approximately express the second term in Eq.~(\ref{eq:global-cons_3})
with $\gamma < 0$ as
\begin{align}
  \sum_{n = 1}^{N} \frac{e^{2g(N-n)}\gamma \rho_{n}}{\Gamma\sin(ka)}
  \sim A' \exp\left[-2\left(\frac{1}{\xi}-g\right)N\right] - e^{2gN}B',
\end{align}
where $A' > 0$ is independent of $N$,
whereas $B' > 0$ may weakly depend on $N$.
Again, $e^{2gN}B'$ should cancel $e^{2g(N-1)}R_{\rm I}$
on the right-hand side of Eq. (\ref{eq:global-cons_3})
to give the exponential decrease in $R_{\rm T}$.
We then find
\begin{align}
  R_{\rm T} \sim A' \exp\left[-2\left(\frac{1}{\xi}-g\right)N\right].
\end{align}
Since the decay length $\xi$ decreases with increasing $|\gamma|$,
this explains the behavior of $R_{\rm T}$ shown in Fig. 5(a).
Figure~5(b) shows $-\langle \log R_{\rm I}\rangle$ for 
$\gamma/\Gamma = 0.0$, $-0.00001$, $-0.0001$, and $-0.001$.
The results shown in this figure are the same as those shown in Fig.~4(b)
because $R_{\rm I}$ in the case of $g = 0.02$ with a given $\gamma$
is identical to that in the case of $g = 0$ with the same $\gamma$
as shown in Appendix.

\section{Concluding remarks}

In a mesoscopic length scale, transport through a conductor is characterized by
the transmission and reflection probabilities,
which are found by solving the corresponding scattering problem.
Here, we have considered a non-Hermitian conductor
connected to two Hermitian leads in a standard two-terminal setup.
As a concrete Hamiltonian describing a non-Hermitian conductor,
we have considered a generic non-Hermitian model
with Hatano--Nelson-type asymmetric hopping
and analyzed the corresponding non-Hermitian scattering problem.
We have seen that the Hermitian interpretation of the scattering problem
based on the transmission and reflection probabilities fails in this case.
Introducing the incoming and outgoing currents
[see Eqs.~(\ref{j_in}) and (\ref{j_out})],
we have proposed the modified continuity equation~(\ref{conti_mod})
with Eq.~(\ref{eq:multiplicative}),
which serves as a local conservation law.
The global probability conservation law~(\ref{glob_cons}) is derived
from this local conservation law.
We have tested the usefulness of our global probability conservation law
in the interpretation of numerical experiments
on non-Hermitian localization and delocalization phenomena.

We have seen that
transport through a non-Hermitian system in the two-terminal setup
is beyond a two-terminal description based on the Landauer formula.
This is because an effective current is either injected from or leaked to
a hypothetical external reservoir assumed to be existing
in the underlying Hermitian model [see Fig.~1(b)].
Such a situation may be more appropriately described by a multiterminal
Landauer formula~\cite{buttiker}
and will deserve a more thorough discussion in a future work.

Related to this, a microscopic derivation of the non-Hermitian effective
Hamiltonian~(\ref{H_S}), or Eq.~(\ref{HN-asymmetric}),
will also deserve a thorough discussion.
In the original paper of Hatano and Nelson,~\cite{HN1}
the effective Hamiltonian~(\ref{HN-asymmetric}) has been introduced to
describe the depinning of a flux line driven by a perpendicular magnetic field
in a cylindrical superconducting shell with columnar pinning centers.
In the present context, Eq.~(\ref{H_S}) is expected to describe
a one-dimensional Hermitian electron system
coupled with an external reservoir [see Fig.~1(b)].
An important problem left for a future study is to determine the circumstances
under which the use of the effective Hamiltonian is fully justified.
A non-Hermitian electron system governed by Eq.~(\ref{HN-asymmetric})
may be realized
if each pair of nearest neighbor sites in the system are coupled with
a reservoir in a particular manner.~\cite{gong,mcdonald,liu}

\section*{Acknowledgment}

This work was supported by JSPS KAKENHI Grant Numbers JP21K03405 and 20K03788.

\section*{Appendix}

In implementing the asymmetry in $\Gamma_{\rm R}$ and $\Gamma_{\rm L}$,
we set $\Gamma_{\rm R} = e^{g}\Gamma$ and
$\Gamma_{\rm L} = e^{-g}\Gamma$,~\cite{HN1,HN2}
which is equivalent to assuming $\Gamma_{\rm R}\Gamma_{\rm L} = \Gamma^{2}$.
Under this setting, it is useful to introduce
the similarity transformation
\begin{align}
      \label{eq:similarity-trans}
   \Lambda = \sum_{n} |n \rangle b_{n} \langle n|
\end{align}
with
\begin{align}
   b_{n} = \left\{ \begin{array}{cc}
                    1 & (n \le 0), \\
                    e^{-g(n-1)} & (1 \le n \le N), \\
                    e^{-g(N-1)} & (N+1 \le n) .
                  \end{array}
           \right.
\end{align}
We can show that $H$ is transformed to $\Lambda H \Lambda^{-1} = H_{g=0}$,
where $H_{g=0}$ represents the Hamiltonian with $\Gamma_{\rm R} = \Gamma_{\rm L} = \Gamma$.
Let us compare the stationary scattering state $|\psi_{g=0}\rangle$
satisfying $H_{g=0}|\psi_{g=0}\rangle = E|\psi_{g=0}\rangle$
and the stationary scattering state $|\psi\rangle$
satisfying $H|\psi\rangle = E|\psi\rangle$.
The relation $\Lambda H \Lambda^{-1} = H_{g=0}$ ensures
\begin{align}
   |\psi_{g=0}\rangle = \Lambda|\psi\rangle ,
\end{align}
which shows that, in the left lead of $n \le 0$,
$|\psi_{g=0}\rangle$ is identical to $|\psi\rangle$ since $b_{n} = 1$.
This means that, for the given $\gamma$ and disorder potential $V_{n}$,
$R_{\rm I}$ associated with $|\psi\rangle$ is identical to
that associated with $|\psi_{g=0}\rangle$.
This statement relies on the similarity transformation;
thus, it does not exactly hold in the case of
$\Gamma_{\rm R}\Gamma_{\rm L} \neq \Gamma^{2}$,
although Eq.~(\ref{eq:global-cons2}) still holds.


\begin{thebibliography}{99}

\bibitem{open_q}
I. Rotter, J. Phys. A: Math. Theor. {\bf 42}, 153001 (2009).

\bibitem{open_q2}
Y. Ashida, Z. Gong, and M. Ueda, Adv. Phys. {\bf 69}, 249 (2020).

\bibitem{opt1}
R. E. Le Levier and D. S. Saxon, Phys. Rev. {\bf 87}, 40 (1952).

\bibitem{opt2}
H. Feshbach, C. E. Porter, and V. F. Weisskopf,
Phys. Rev. {\bf 96}, 448 (1954).

\bibitem{opt3} P. E. Hodgson, Rep. Prog. Phys. {\bf 47}, 613 (1984).

\bibitem{andrianov} A.A. Andrianov, M.V. Ioffe, F. Cannata, and J.-P. Dedonder,
Int. J. Mod. Phys. A {\bf 14}, 2675 (1999).

\bibitem{levani} G. L\'{e}vai, F. Cannata, and A. Ventura
J. Phys. A {\bf 34},  839 (2001).

\bibitem{deb} R. N. Deb, A. Khare, and B. D. Roy,
Phys. Lett. A {\bf 307}, 215  (2003).

\bibitem{muga} J. G. Muga, J. P. Palao, B. Navarro, and I. L. Egusquiza,
Phys. Rep. {\bf 395}, 357 (2004).

\bibitem{cannata} F. Cannata, J.-P. Dedonder, and A. Ventura,
Ann. Phys. {\bf 322}, 397 (2007).

\bibitem{jones} H. F. Jones, Phys. Rev. D {\bf 76}, 125003 (2007).

\bibitem{znojil} M. Znojil, Phys. Rev. D {\bf 78}, 025026 (2008).

\bibitem{reso1} N. Hatano,  K. Sasada,  H. Nakamura,  and T. Petrosky,
Prog. Theor. Phys. {\bf 119}, 187 (2008).

\bibitem{jin} L. Jin and Z. Song, Phys. Rev. A {\bf 8}5, 012111 (2012).

\bibitem{abhinav} K. Abhinav, A. Jayannavar, and P. K. Panigrahi,
Ann. Phys. {\bf 331}, 110 (2013).

\bibitem{reso2} N. Hatano, Fortschr. Phys. {\bf 61}, 238 (2013).

\bibitem{kalozoumis} P. A. Kalozoumis, G. Pappas, F. K. Diakonos,
and P. Schmelcher, Phys. Rev. A {\bf 90}, 043809 (2014).

\bibitem{garmon} S. Garmon, M. Gianfreda, and N. Hatano,
Phys. Rev. A {\bf 92}, 022125 (2015).

\bibitem{zhu} B. Zhu, R. L\"{u}, and S. Chen,
Phys. Rev. A {\bf 93}, 032129 (2016).

\bibitem{ruschhaupt}  A. Ruschhaupt, T. Dowdall, M. A. Sim\'{o}n,
and J. G. Muga, EPL {\bf 120} 20001 (2017).

\bibitem{burke} P. C. Burke, J. Wiersig, and M. Haque,
Phys. Rev. A {\bf 102}, 012212 (2020).

\bibitem{shobe} K. Shobe, K. Kuramoto, K.-I. Imura, and N. Hatano,
Phys. Rev. Research {\bf 3}, 013223 (2021).

\bibitem{comment1} If $\psi_n(t)$ is not normalizable as in the case of
a stationary scattering state,
$\rho_n(t)=|\psi_n(t)|^2$ cannot be simply interpreted as the probability
of finding an electron at the $n$th site.
However, if many stationary scattering states are superposed to obtain
a wave packet $\Psi_n(t)$ that satisfies the normalization condition
$\sum_{n}|\Psi_n(t)|^{2} = 1$, the probabilistic interpretation
is applicable to $\rho_n(t)=|\Psi_n(t)|^2$ in a rigorous sense.

\bibitem{bagchi} B. Bagchi, C. Quesne, and M. Znojil,
Mod. Phys. Lett. A {\bf 16}, 2047 (2001).

\bibitem{NH2019TD} K. Shobe,
``Spontaneous current, skin and proximity effects in non-Hermitian systems'', 
presented at IIS-Chiba Workshop on Non-Hermitian Quantum Mechanics (NH2019TD),
2019.

\bibitem{HN1} N. Hatano and D. R. Nelson, Phys. Rev. Lett. {\bf 77}, 570 (1996).
\bibitem{HN2} N. Hatano and D. R. Nelson, Phys. Rev. B {\bf 56}, 8651 (1997).

\bibitem{KK} An alternative recipe is given in
K. Kawabata, T. Numasawa, and S. Ryu, Phys. Rev. X 13, 021007 (2023).
This recipe uses Eq.~(\ref{j0}) with $\Gamma$ replaced with
$(\Gamma_{\rm R}+\Gamma_{\rm L})/2$ as an expression of probability current
and therefore essentially differs from that we propose in this paper
[see Eqs.~(\ref{conti_mod})--(\ref{j_out})].

\bibitem{gong}  Z. P. Gong, Y. Ashida, K. Kawabata, K. Takasan, S.
Higashikawa, and M. Ueda, Phys. Rev. X {\bf 8}, 031079 (2018).

\bibitem{mcdonald} A. McDonald, R. Hanai, and A. A. Clerk,
Phys. Rev. B {\bf 105}, 064302 (2022).

\bibitem{liu} P. Liu, X. Y. Zhang, X. Z. Hao, Y. H. Zhou, S. C. Hou,
and X. X. Yi, Phys. Rev. A {\bf 107}, 053515 (2023).

\bibitem{HN_exp1} M. Brandenbourger, X. Locsin, E. Lerner, and C. Coulais,
Nat. Commun. {\bf 10}, 4608 (2019).
 
\bibitem{HN_exp2} S. Weidemann, M. Kremer, T. Helbig, T. Hofmann,
A. Stegmaier, M. Greiter, R. Thomale, and A. Szameit,
Science {\bf 368}, 311 (2020).

\bibitem{HN_exp3} T. Helbig, T. Hofmann, S. Imhof, M. Abdelghany,
T. Kiessling, L. W. Molenkamp, C. H. Lee, A. Szameit, M. Greiter,
and R. Thomale, Nat. Phys. {\bf 16}, 747 (2020).

\bibitem{HN_exp4} L. S. Palacios, S. Tchoumakov, M. Guix, I. Pagonabarraga,
S. S\'{a}nchez, and A. G. Grushin, Nat. Commun. {\bf 12}, 4691 (2021).

\bibitem{landauer} R. Landauer, Philos. Mag. {\bf 21}, 863 (1970).

\bibitem{buttiker} M. B\"{u}ttiker, Y. Imry, R. Landauer, and S. Pinhas,
Phys. Rev. B {\bf 31}, 6207 (1985).


\end{thebibliography}
\end{document}